\documentclass[reprint,aps,prl,superscriptaddress,floatfix]{revtex4-2}

\usepackage{pdfpages} 
\usepackage{pgffor} 
\makeatletter
\AtBeginDocument{\let\LS@rot\@undefined}
\makeatother
\def\supplementfilename{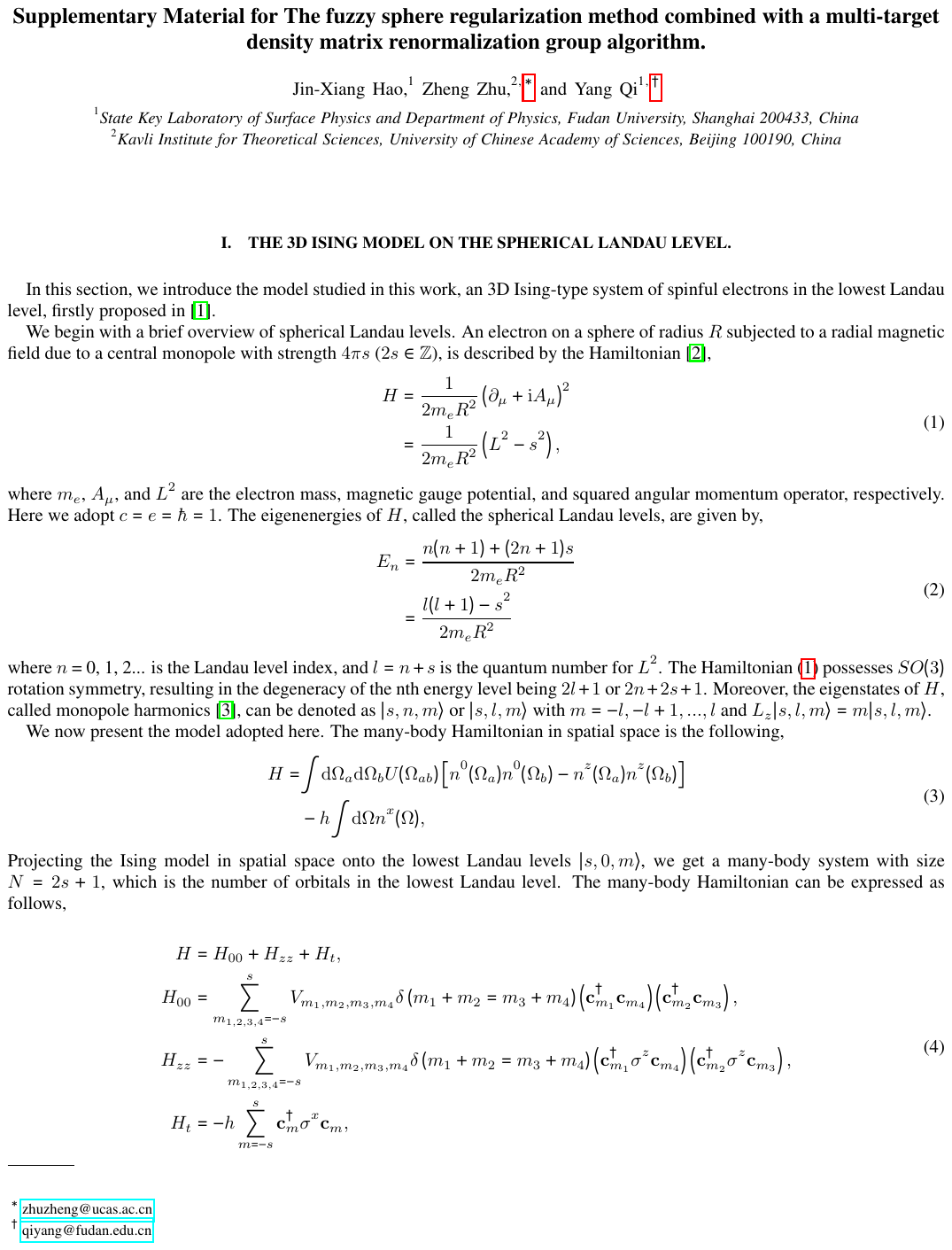}
\pdfximage{\supplementfilename}
\def\numbersupplementpages{\the\pdflastximagepages}
\newif\ifarXiv
\arXivtrue

\usepackage{amsmath}
\usepackage{amssymb}
\usepackage{mdsymbol}
\usepackage{graphicx}
\usepackage{dsfont}
\usepackage{braket}
\usepackage{hyperref}
\usepackage{amsfonts}

\usepackage{textcomp}
\usepackage{times}
\usepackage{float}
\usepackage{color}
\usepackage{lipsum}
\usepackage{multirow}
\begin{document}

\title{Multi-target density matrix renormalization group for 3D CFTs on the fuzzy sphere}

\author{Jin-Xiang Hao}
\affiliation{State Key Laboratory of Surface Physics and Department of Physics, Fudan University, Shanghai 200433, China}

\author{Zheng Zhu}
\email{zhuzheng@ucas.ac.cn}
\affiliation{Kavli Institute for Theoretical Sciences, University of Chinese Academy of Sciences, Beijing 100190, China}

\author{Yang Qi}
\email{qiyang@fudan.edu.cn}
\affiliation{State Key Laboratory of Surface Physics and Department of Physics, Fudan University, Shanghai 200433, China}

\begin{abstract}
The fuzzy sphere regularization provides a powerful framework for studying three-dimensional (3D) conformal field theories (CFTs) by mapping them onto 
numerically tractable lattice models on the spherical lowest Landau level. However, the system sizes accessible to this method have been limited by the exact diagonalization (ED). 
In this work, we transcend this limitation by combining the fuzzy sphere regularization with a sophisticated multi-target density matrix renormalization group (DMRG) algorithm. 
Focusing on the 3D Ising-type model on the spherical lowest Landau level, we calculate the 24 low-lying energies at a larger system size than 
previously feasible with ED. At criticality, we extract the scaling dimensions of six primary operators, and the results show 
significantly improved agreement with bootstrap benchmarks compared to previous ED results at smaller sizes. Our approach allows us to efficiently 
target multiple excited states in larger systems beyond the reach of exact diagonalization. This study establishes the fuzzy sphere regularization combined with advanced DMRG techniques as a 
powerful and general framework for precision physics in 3D CFTs. 
\end{abstract}

\maketitle

\paragraph{Introduction.}
Conformal field theories (CFTs) play a foundational role in our understanding of quantum critical phenomena in quantum many-body systems and quantum gravity in high-energy physics \cite{polyakov1970conformal, belavin1984infinite, cardy1984conformal, francesco2012conformal, maldacena1999large}. 
Despite its importance, it is highly challenging, especially for 3D CFTs \cite{francesco2012conformal, brower2013lattice, brower2021radial}, to extract precise conformal data such as the scaling dimensions of primary operators. 
Recently, a numerical framework known as the fuzzy sphere regularization has been introduced \cite{zhu2023uncovering}, offering a powerful method for realizing 
3D CFTs in finite and numerically tractable systems \cite{han2024conformal, dey2025conformal, he2025free, taylor2025conformal, voinea2025critical, zhou2025free, zhou2023mathrm, xstj-xvcy, yang2025conformal, zhou2025chern, cruz2025yang, fan2025simulating, miro2025flowing, hu2023operator, han2023conformal, hu2024entropic, fan2024note, fardelli2025constructing, hu2024solving, zhou2024g, cuomo2024impurities, zhou2025studying, dedushenko2024ising}. This framework constructs a lattice model by projecting 3D Ising models in spatial space onto 
the lowest Landau level (LLL) of a spherical geometry \cite{haldane1983fractional}, and calculates the energy spectrum utilizing exact diagonalization (ED). From the eigenenergies of the lattice model, 
the scaling dimensions of the corresponding conformal operators can be obtained via the state-operator correspondence.

Nonetheless, the fuzzy sphere approach has been limited by the system sizes accessible to ED, which is constrained by the exponential 
growth of the Hilbert space, the so-called exponential wall. While ED has provided valuable initial results for small systems, its inability to 
handle larger sizes restricts the accuracy of the extracted CFT data. There are some works overcoming this limitation by using density matrix renormalization group (DMRG) techniques \cite{hu2023operator, yang2025microscopic, hu2024solving, zhou2024g, han2024conformal, dey2025conformal, taylor2025conformal, chen2024emergent, zhou2025studying, wiese2025localizing, 10.21468/SciPostPhys.19.3.076}, 
but they mainly focus on the ground state and relatively small numbers of excited states. As a result, the reliable identification of more primary operators from the low-energy spectrum remains practically limited.

In this work, we overcome these challenges by combining the fuzzy sphere regularization with a multi-target DMRG 
algorithm, enhanced by the use of block Lanczos incorporating the symmetries of the model.
To demonstrate our approach, we focus on the 3D Ising-type model defined on LLLs, which is proposed in Ref.~\cite{zhu2023uncovering} (see Supplemental Material for a brief overview).
By targeting four low-lying states in each symmetry sector, we are able to extract the scaling dimension of six primary operators and 
organize the low-lying excitations into conformal multiplets.
Our DMRG results, obtained at a larger system size ($s = 15.5$, $N = 32$ orbitals),
show significantly improved agreement with bootstrap benchmarks compared to previous ED results at a smaller size ($s = 7.5$, $N = 16$ orbitals), demonstrating the 
efficacy of our method.

This study not only advances the numerical determination of 3D Ising CFT data but also establishes the fuzzy sphere regularization combined with advanced DMRG techniques as a powerful and general framework for future studies of more complex CFTs, particularly for models beyond the reach of ED, such as multi-component (orbital, layer, band, spin, and beyond) systems. 

The rest of this paper is organized as follows. We first describe the multi-targeted DMRG algorithm with block Lanczos and analyze the symmetries used in our numerical calculations of the 3D Ising-type model. We then present our numerical results, including the low-energy spectrum across different symmetry sectors, the determination of the critical point for the paramagnetic-ferromagnetic phase transition, and the extraction of scaling dimensions for primary operators and their descendants. We also discuss the computational cost of this DMRG algorithm and its potential applications.
A brief review of the 3D Ising model we use and the form of the squared
angular momentum operator $\vec L^2$ under Landau levels is included in the Supplemental Material (SM)\footnote{See Supplemental Material for additional details on the 3D Ising model and the representation of the total squared angular momentum operator under Landau levels, which includes Refs.~\cite{haldane1983fractional,wu1976dirac}}.

\paragraph{Methods.}
The DMRG is an efficient method for studying one-dimensional and quasi-one-dimensional quantum systems \cite{white1992density}, and has achieved remarkable success in computing ground-state properties. 
For excited states, the standard DMRG strategy introduces iteratively penalty terms into the original Hamiltonian $H_0$ \cite{annurev:/content/journals/10.1146/annurev-conmatphys-020911-125018}. In this approach,
the ground state $|\psi_0\rangle$ of $H_0$ is first obtained, and the Hamiltonian is then modified as $H^\prime=H_0+w\left| \psi_0 \rangle \langle \psi_0 \right|$, where a real positive number $w$ raises the energy of $|\psi_0\rangle$ by $w$, enabling the variational minimum of $H^\prime$ to correspond to the 
first excited state of $H_0$. This process can be repeated by penalizing additional states to access higher excitations. However, due to the presence of truncation errors in the calculations, the obtained states are not exact eigenstates of $H_0$. This results in a projection matrix that does not strictly commute with $H_0$, making the approach less reliable for targeting multiple excited states. Furthermore, the computational cost increases significantly as the number of target states grows.

An alternative approach for accessing excited states is the multi-targeted DMRG algorithm with block Lanczos \cite{baker2024direct}. 
This method comprises two key ingredients: first, generalization of the matrix product state (MPS) to a bundled MPS, which incorporates an 
extra index in the orthogonality center to label distinct excitations; second, the replacement of the standard Lanczos step in DMRG with a block Lanczos procedure, which matches 
updating the orthogonality center of a bundled MPS. Using this approach, multiple excited states can be obtained accurately without too much computational cost. 

In our numerical simulations, symmetries not only significantly reduce the computational cost of DMRG, but also enable access to the excited states of the Hamiltonian. 
By isolating symmetry sectors, the excited states of interest can be found by searching for the lowest-energy state within each sector \cite{white1992density}. 
The Hamiltonian studied in this work has $SO(3)$ sphere rotation symmetry and $\mathbb{Z}_2$ Ising symmetry \cite{zhu2023uncovering}. 
Therefore, we can leverage rotational symmetry about the $z$-axis $U(1)$ (subgroup of $SO(3)$) 
and $\mathbb{Z}_2$ symmetries to handle the model in our DMRG calculation. 
To directly utilize $\mathbb{Z}_2$ symmetry, we need to apply a $\mathbb{Z}_2$ unitary transformation to the model, considering that the Hamiltonian (14) formulated in Ref.~\cite{zhu2023uncovering} has spinful degrees of freedom, 
which is not an irreducible representation of the $\mathbb{Z}_2$ transformation. 
Specifically, we introduce a unitary transformation,
\begin{equation}
  \label{202502222135}
  \begin{aligned}
    &c_\uparrow=\frac{1}{\sqrt2}\left(c_+ + c_-\right), \quad c_\downarrow=\frac{1}{\sqrt2}\left(c_+ - c_-\right),
\end{aligned}
\end{equation}
to map the model from the spin basis $\ket{\uparrow}$ and $\ket{\downarrow}$ to the basis of irreducible representations of the $\mathbb{Z}_2$ symmetry $\ket{+}$ and $\ket{-}$ (or equivalently, exchanging the $X$ and $Z$ Pauli matrices in the model). Applying this transformation to the model, 
we get the following form,
\begin{equation*} 
\label{202502201343} 
\begin{aligned} 
H_0&=\frac{1}{4}\sum_{m_1\lambda_1}\sum_{m_2\lambda_2} \sum_{m_3\lambda_3}\sum_{m_4\lambda_4} \left(\lambda_2\lambda_3 + \lambda_1\lambda_4\right) \delta_{m_1+m_2,\,m_3+m_4} 
\\ &\quad V_{m_1m_2m_3m_4}c^{\dagger}_{m_1\lambda_1}c^{\dagger}_{m_2\lambda_2}c_{m_3\lambda_3}c_{m_4\lambda_4}-h\sum_{m\lambda}\lambda c^{\dagger}_{m\lambda} c_{m\lambda}, 
\end{aligned} 
\end{equation*}
where $\lambda=\pm1$ and $m=-s,...,s$. The U(1) symmetry corresponds to the conservation of the $z$‑component angular momentum $m_z$. It is important to note that a given $m_z>0$ value can 
be associated with multiple angular momentum quantum numbers $l>m_z$. For example, states with $m_z=0$ may have $l=0, 1, 2, ...$, i.e., $l$ can take 
any non‑negative integer. To selectively target states with specific angular momentum quantum numbers $l=m_z$ 
for a given $m_z>0$, we add a squared angular momentum operator $\vec{L}^2$ to the Hamiltonian $H_0$,
\begin{equation} 
\label{202510040012} 
H=H_0+\lambda\vec{L}^2. 
\end{equation}
where $\lambda$ is a positive real number. This term penalizes states with $l>m_z$ by raising their energies, thereby suppressing them during the variational optimization in DMRG. The representation of the total squared angular momentum operator $\vec{L}^2$ in the Landau levels basis is 
derived in Sec. II of the SM. The Hamiltonian in Eq.~\eqref{202510040012} is the model studied in the following sections.

\begin{table}[tbp]
\centering
\begin{tabular}{|c|c|c|c|c|c|}
\hline
\multirow{2}{*}{$\mathbb Z_2$} & \multirow{2}{*}{$l$} & \multicolumn{4}{|c|}{\textbf{operators}} \\
\cline{3-6}
& & 0 & 1 & 2 & 3\\
\hline
\multirow{3}{*}{1} & 0 & $I$ & $\epsilon$ & $\Box\epsilon$ & $\epsilon^\prime$ \\
                     \cline{2-6}               
                   & 1 & $\partial_\mu\epsilon$ & $\partial_\mu\Box\epsilon$ & $\partial_\mu\epsilon^\prime$ & $\partial_\mu\Box^2\epsilon$ \\    
                     \cline{2-6} 
                   & 2 & $T_{\mu\nu}$ & $\partial_\mu\partial_\nu\epsilon$ & $\epsilon_{\nu\rho\mu}\partial_\rho T_{\mu\nu}$ & $\Box T_{\mu\nu}$ \\                                                            
\hline
\multirow{3}{*}{-1} & 0 & $\sigma$ & $\Box\sigma$ & $\Box^2\sigma$ & $\sigma^\prime$ \\ 
                      \cline{2-6}
                    & 1 & $\partial_\mu\sigma$ & $\partial_\mu\Box\sigma$ & $\partial_{\mu}\sigma_{\mu\nu}$ & $\partial_\mu\Box^2\sigma$ \\  
                      \cline{2-6}
                    & 2 & $\partial_{\mu}\partial_{\nu}\sigma$ & $\sigma_{\mu\nu}$ & $\Box\partial_\mu\partial_\nu\sigma$ & $\epsilon_{\nu\rho\mu}\partial_\rho\sigma_{\mu\nu}$ \\ 
\hline
\end{tabular}
\caption{The operators corresponding to the four lowest-energy states in each symmetry sector $\left(\mathbb{Z}_2, l\right)$}.
\label{20251017}
\end{table}
\paragraph{Results.}
In this section, we compute the low-energy spectrum of the Hamiltonian in Eq.~\eqref{202510040012} using a multi-targeted DMRG algorithm with block Lanczos \cite{ITensor, baker2024direct}. The $U(1)$ and $\mathbb{Z}_2$ symmetries described above are implemented to extract the corresponding Ising-CFT information.  We target the four lowest energy levels in each of the symmetry sectors $(z = \pm 1; m_z= 0, 1, 2)$ in the finite-size truncation $s=15.5$ (i.e., $N=32$ orbitals) with maximum bond dimension $D=10000$ and truncation errors below $10^{-8}$. The lowest state in the sector $(z=1, m_z=0)$ corresponds to the ground state.

We first rescale the energy gap (the difference between the excited-state energy and the ground-state energy) of each excited state.
According to the state-operator correspondence \cite{zhu2023uncovering}, the scaling dimension 
of operators is proportional to the energy of the corresponding state, and the proportional coefficient is size-dependent. Therefore, we locate the lowest 
energy level in the sector $(z=1, m_z=2)$, which corresponds to the energy momentum tensor $T_{\mu\nu}$ with scaling dimension $\Delta_{T_{\mu\nu}}=3$, and rescale energy gaps of all states such that the energy gap of the lowest state in the sector $(z=1, m_z=2)$ equals 3.

\begin{table*}[tpb]
\centering
\resizebox{0.55\textwidth}{!}{%
\begin{tabular}{c|c|c|c|c|c|c}
\hline
\textbf{primary operators} & $\sigma$ & $\epsilon$ & $\epsilon^{\prime}$ & $\sigma^{\prime}$ & $T_{\mu\nu}$ & $\sigma_{\mu\nu}$\\
\hline
\textbf{symmetry sectors $(z,l)$} & $(-1,0)$ & $(1,0)$ & $(1,0)$ & $(-1,0)$ & $(1,2)$ & $(-1,2)$\\
\hline
\textbf{Bootstrap} & 0.518 & 1.413 & 3.830 & 5.291 & 3.00 & 4.180\\
\hline
\textbf{Fuzzy sphere $s=15.5$} & 0.519 & 1.413 & 3.836 & 5.299 & 3.00 & 4.205\\
\hline
\textbf{Errors$_{15.5}$} & 0.19\% & 0.00\% & 0.16\% & 0.15\% & / & 0.60\%\\
\hline
\textbf{Fuzzy sphere $s=7.5$} & 0.524 & 1.414 & 3.838 & 5.303 & 3.00 & 4.214\\
\hline
\textbf{Errors$_{7.5}$} & 1.16\% & 0.07\% & 0.21\% & 0.23\% & / & 0.81\%\\
\hline
\end{tabular}%
}
\caption{Scaling dimension of six Ising-CFT primary operators obtained by the fuzzy sphere regularization with the truncation $s=15.5$ using multi-target DMRG, compared with results from exact diagonalization at $s=7.5$~\cite{zhu2023uncovering} and conformal bootstrap~\cite{David2017bootstrap}.}
\label{scaling dimension}
\end{table*}

\begin{figure*}[tbp]
  \centering
  \includegraphics[width=0.7\textwidth]{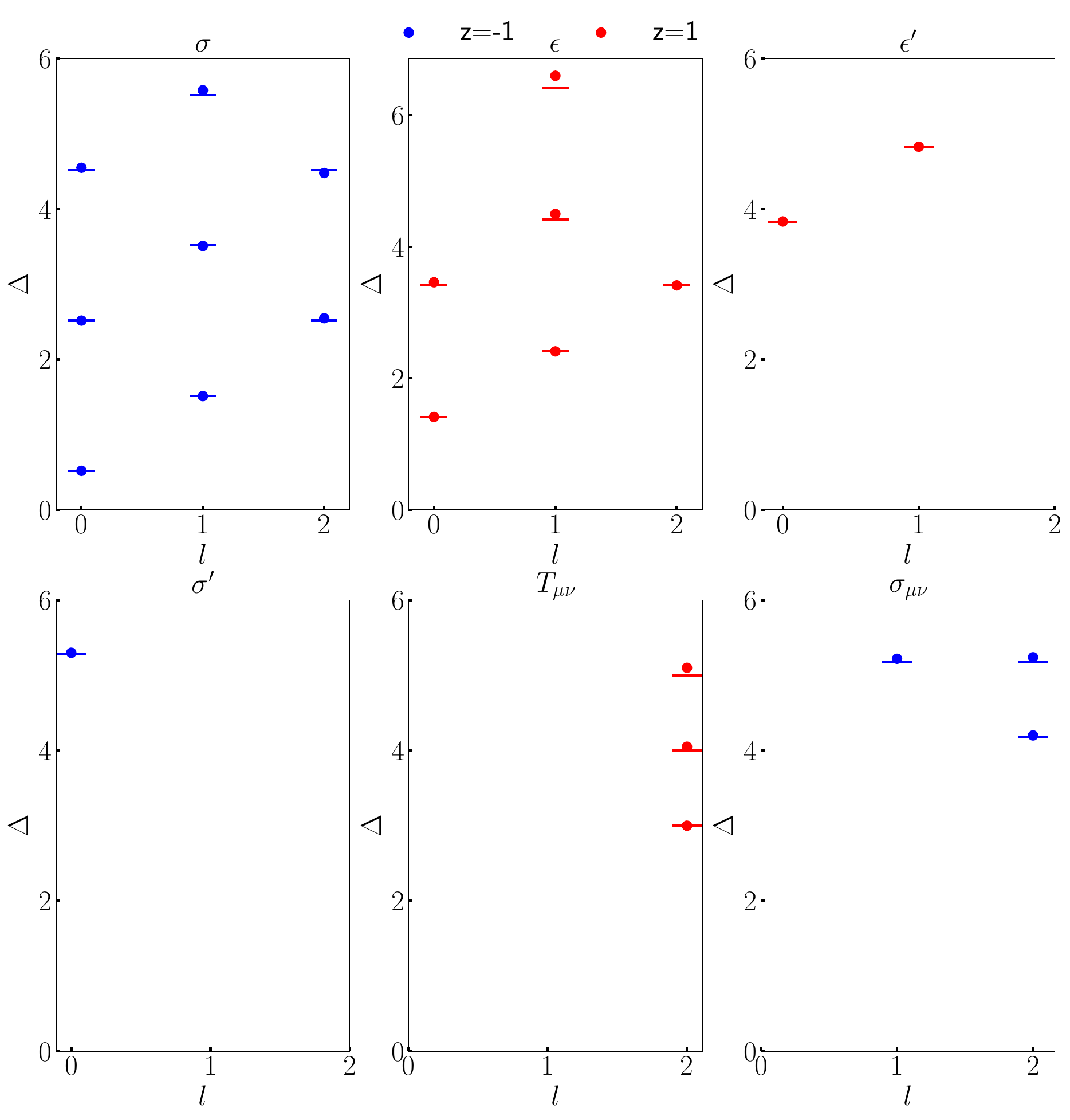}
  \caption{Conformal multiplet of several primary operators. Dots show the results from the fuzzy sphere regularization with the truncation $s=15.5$ (corresponding to $N=32$ orbitals) via multi-target DMRG calculations, and lines represent corresponding bootstrap benchmarks.}
  \label{tower}
\end{figure*}

Next, we determine the critical point $(U_{0c},h_c)$ for the paramagnetic-ferromagnetic phase transition which the system undergoes as we tune $h$ and $U_0$. 
The standard practice for locating the critical point is to analyze the crossing of the order parameters or Binder cumulants across different system sizes. However, for the model defined on the Landau level, 
their energy spectrum exhibits a characteristic conformal tower structure at the critical point \cite{zhu2023uncovering}. Thus, examining this 
special energy-level structure is a more accurate approach to identifying the critical point in finite-size systems. For the model Hamiltonian in Eq.~\eqref{202510040012}, the lowest energy $e_{0}$ in the sector 
$(z=-1, m_z=0)$ corresponds to the order parameter operator $\sigma$, while the lowest energy $e_{1}$ in the sector $(z=-1, m_z=1)$ corresponds to 
its descendant $\partial_\mu \sigma$, and lowest energy $e_{2}$ in the sector $(z=-1, m_z=2)$ corresponds to $\partial_\mu \partial_\nu \sigma$ (notice that here $e_{0}$, $e_{1}$, and $e_{2}$ are rescaled energy gaps). By monitoring the values of $e_0$, $e_{1}$, and $e_{2}$ across different parameters $h$ and $U_0$, we identify the critical point $(U_0=4.75, h_c=3.153)$ as the condition $e_0+1=e_{1}$ and $e_{1}+1=e_{2}$ is approximately satisfied.
We note that the location of the critical point is refined at the largest system size we simulate ($s=15.5$), as the location drifts with the increasing of system size.

At the critical point $(U_0=4.75, h_c=3.153)$, we extract several primary operators and their descendants (i.e. conformal multiplet) from the rescaled energy gaps, as summarized in  Table.~\ref{20251017}. 
Among the 24 computed energies (the four lowest energy levels in each of the symmetry sectors $(z = \pm 1; m_z= 0, 1, 2)$), we identify 6 primary operators 
and obtain their scaling dimension as shown in Table.~\ref{scaling dimension}. In the table, we additionally list the results using exact diagonalization 
with the truncation $s=7.5$ \cite{zhu2023uncovering}, and bootstrap results \cite{David2017bootstrap} for comparative analysis. Furthermore, 
the 24 energies are classified according to their correspondence with primary operators and the associated descendant operators by conformal tower relations, 
as shown in Fig.~\ref{tower}.

\paragraph{Summary and Outlook.}
In this work, we propose a general numerical framework based on a multi-target DMRG algorithm with block Lanczos that overcomes the system-size limitations inherent to exact diagonalization in fuzzy-sphere regularized conformal field theories. This allows access to significantly larger system sizes and efficient targeting of multiple low-energy excitations, enabling the systematic extraction of conformal data.
As a concrete demonstration, we apply this approach to the 3D Ising-CFT model constructed by the fuzzy sphere regularization. At criticality, the extracted scaling dimensions of Ising CFT primary operators show substantially improved agreement with bootstrap benchmarks compared to exact diagonalization results for smaller systems, highlighting both the accuracy and robustness of the numerical method.

Our results demonstrate that, for models with fuzzy sphere regularization, the multi-target DMRG algorithm provides an efficient and scalable approach to extracting conformal data. While exact diagonalization is severely limited by the exponential growth of the Hilbert space, DMRG can access substantially larger systems. By decomposing the Hilbert space into symmetry sectors, multiple excited states can be targeted systematically within each sector, enabling the extraction of scaling dimensions for primary operators and their descendants. This framework thus opens the door to investigating more complex conformal field theories, such as multi-component (orbital, layer, band, spin, and beyond) systems that are beyond the reach of exact diagonalization on the fuzzy sphere.

Finally, we note that the computational cost of the DMRG method exhibits a $\sqrt N$ scaling with the system size $N$, which originates from the geometry of the fuzzy-sphere construction. The entanglement entropy follows the area law, where the relevant boundary is the circumference of the bipartition, proportional to the sphere radius $R$. Since the surface area of the sphere is proportional to the monopole strength $s$, and $s$ is proportional to the number of orbitals $N$, we have 
$N \propto R^2$ or equivalently $R \propto \sqrt{N}$. As a consequence, the entanglement entropy grows as $\sqrt{N}$, and ultimately dictates $\sqrt{N}$ scaling of the DMRG bond dimension and computational cost. 
While our present implementation only uses the Abelian $U(1)$ symmetry associated with the conservation of the $z$-component of angular momentum, an important future direction is to incorporate the full non-Abelian $SO(3)$ rotational symmetry. Such an extension is expected to further improve numerical efficiency and pave the way toward accessing substantially larger systems within this framework.

\begin{acknowledgments}
  The authors thank Wei Zhu and Shuai Yang for invaluable discussions.
  This work was supported by National Key R\&D Program of China (Grant No. 2022YFA1403402), by National Natural Science Foundation of China (Grant Nos. 12174068, 92477106), by the Science and Technology Commission of Shanghai Municipality (Grant Nos. 24LZ1400100 and 23JC1400600), and by the Fundamental Research Funds for the Central Universities.
  Y.Q. acknowledges the support of Shuguang Program of Shanghai Education Development Foundation and Shanghai Municipal Education Commission.
\end{acknowledgments}

\bibliography{ref}

\ifarXiv
    

\section*{Supplementary material}

# Supplementary Material for The fuzzy sphere regularization method combined with a multi-target density matrix renormalization group algorithm.

Jin-Xiang Hao,[1] Zheng Zhu,[2,*] and Yang Qi[1,†]

[1]*State Key Laboratory of Surface Physics and Department of Physics, Fudan University, Shanghai 200433, China*
[2]*Kavli Institute for Theoretical Sciences, University of Chinese Academy of Sciences, Beijing 100190, China*

## I. THE 3D ISING MODEL ON THE SPHERICAL LANDAU LEVEL.

In this section, we introduce the model studied in this work, an 3D Ising-type system of spinful electrons in the lowest Landau level, firstly proposed in [1].

We begin with a brief overview of spherical Landau levels. An electron on a sphere of radius $R$ subjected to a radial magnetic field due to a central monopole with strength $4\pi s$ ($2s \in \mathbb{Z}$), is described by the Hamiltonian [2],

$$H = \frac{1}{2m_e R^2}\left(\partial_\mu + \mathrm{i} A_\mu\right)^2 \\ = \frac{1}{2m_e R^2}\left(L^2 - s^2\right), \quad (1)$$

where $m_e$, $A_\mu$, and $L^2$ are the electron mass, magnetic gauge potential, and squared angular momentum operator, respectively. Here we adopt $c = e = \hbar = 1$. The eigenenergies of $H$, called the spherical Landau levels, are given by,

$$E_n = \frac{n(n+1) + (2n+1)s}{2m_e R^2} \\ = \frac{l(l+1) - s^2}{2m_e R^2} \quad (2)$$

where $n = 0, 1, 2...$ is the Landau level index, and $l = n + s$ is the quantum number for $L^2$. The Hamiltonian (1) possesses $SO(3)$ rotation symmetry, resulting in the degeneracy of the nth energy level being $2l+1$ or $2n+2s+1$. Moreover, the eigenstates of $H$, called monopole harmonics [3], can be denoted as $|s, n, m\rangle$ or $|s, l, m\rangle$ with $m = -l, -l+1, ..., l$ and $L_z|s, l, m\rangle = m|s, l, m\rangle$.

We now present the model adopted here. The many-body Hamiltonian in spatial space is the following,

$$H = \int \mathrm{d}\Omega_a \mathrm{d}\Omega_b U(\Omega_{ab}) \left[n^0(\Omega_a)n^0(\Omega_b) - n^z(\Omega_a)n^z(\Omega_b)\right] \\ - h \int \mathrm{d}\Omega n^x(\Omega), \quad (3)$$

Projecting the Ising model in spatial space onto the lowest Landau levels $|s, 0, m\rangle$, we get a many-body system with size $N = 2s + 1$, which is the number of orbitals in the lowest Landau level. The many-body Hamiltonian can be expressed as follows,

$$H = H_{00} + H_{zz} + H_t, \\ H_{00} = \sum_{m_{1,2,3,4}=-s}^{s} V_{m_1,m_2,m_3,m_4} \delta(m_1 + m_2 = m_3 + m_4)\left(\mathbf{c}^\dagger_{m_1}\mathbf{c}_{m_4}\right)\left(\mathbf{c}^\dagger_{m_2}\mathbf{c}_{m_3}\right), \\ H_{zz} = -\sum_{m_{1,2,3,4}=-s}^{s} V_{m_1,m_2,m_3,m_4} \delta(m_1 + m_2 = m_3 + m_4)\left(\mathbf{c}^\dagger_{m_1}\sigma^z\mathbf{c}_{m_4}\right)\left(\mathbf{c}^\dagger_{m_2}\sigma^z\mathbf{c}_{m_3}\right), \\ H_t = -h \sum_{m=-s}^{s} \mathbf{c}^\dagger_m \sigma^x \mathbf{c}_m, \quad (4)$$

---

* zhuzheng@ucas.ac.cn
† qiyang@fudan.edu.cn

where $\mathbf{c}_m = (c_{m,\uparrow}, c_{m,\downarrow})^T$ is the fermionic annihilation operator of an electron on the lowest Landau level $|s, 0, m\rangle$ with spin up or down, $\sigma^{x,z}$ are the Pauli matrices acting on the spin space, $h$ is the transverse field strength, and $V_{m_1,m_2,m_3,m_4}$ is the interaction matrix element defined as,

$$V_{m_1,m_2,m_3,m_4} = U_0 (4s+1) \begin{pmatrix} s & s & 2s \\ m_1 & m_2 & -m_1-m_2 \end{pmatrix} \begin{pmatrix} s & s & 2s \\ m_4 & m_3 & -m_3-m_4 \end{pmatrix} \\ + (4s-1) \begin{pmatrix} s & s & 2s-1 \\ m_1 & m_2 & -m_1-m_2 \end{pmatrix} \begin{pmatrix} s & s & 2s-1 \\ m_4 & m_3 & -m_3-m_4 \end{pmatrix}, \tag{5}$$

where $\begin{pmatrix} j_1 & j_2 & j_3 \\ m_1 & m_2 & m_3 \end{pmatrix}$ is the Wigner 3j-symbol, and $U_0$ is the interaction strength. Variations in $h$ and $U_0$ can drive the system through a quantum phase transition between a paramagnetic phase and a ferromagnetic phase [1]. Moreover, the Hamiltonian (4) has three symmetries, $SO(3)$ sphere rotation symmetry, $\mathbb{Z}_2$ Ising symmetry, and parity symmetry.

## II. THE REPRESENTATION OF THE TOTAL SQUARED ANGULAR MOMENTUM OPERATOR UNDER LANDAU LEVELS.

In this section, we discuss the representation of the total squared angular momentum operator $L^2$ in the Landau levels basis. The total squared angular momentum operator $L^2$ with system size $N$ can be expressed as,

$$L^2 = \left( \sum_{i=1}^{N} \vec{L}_i \right)^2 \\ = 2 \cdot \sum_{i<j} \vec{L}_i \cdot \vec{L}_j + \sum_{i=1}^{N} \vec{L}_i^2 \\ = 2 \cdot \sum_{i<j} \left( L_i^z \cdot L_j^z + \frac{1}{2}(L_i^+ L_j^- + L_j^+ L_i^-) \right) + \sum_{i=1}^{N} \vec{L}_i^2, \tag{6}$$

where $L_i^z$ is the z-component of the angular momentum operator of the ith particle, and $L_i^+$ ($L_i^-$) is the raising (lowering) operator of the ith particle. Considering the action of the angular momentum operator on the lowest Landau levels with monopole strength $s$ is given by,

$$\vec{L}^2 |s, 0, m\rangle = s(s+1)|s, 0, m\rangle, \tag{7}$$

and,

$$L^z |s, 0, m\rangle = m|s, 0, m\rangle, \tag{8}$$

the equation (6) can be written as follows,

$$\left( \sum_{i=1}^{N} \vec{L}_i \right)^2 = 2 \cdot \sum_{i<j} \left( L_i^z \cdot L_j^z + \frac{1}{2}(L_i^+ L_j^- + L_j^+ L_i^-) \right) + \sum_{i=1}^{N} \vec{L}_i^2 \\ = \sum_{i<j} \left( 2 m_i^z m_j^z \cdot \hat{N}_i \hat{N}_j + L_i^+ L_j^- + L_j^+ L_i^- \right) + \sum_{i=1}^{N} s(s+1) \cdot \hat{N}_i, \tag{9}$$

where $\hat{N}_i$ is the number operator of the ith particle.

Next, we derive the form of the off-diagonal term $L_i^+ L_j^- + L_j^+ L_i^-$ in (9). For simplicity, we temporarily consider the two-particle case. For the two-particle coupled angular momentum state,

$$|m_1, m_2\rangle = \frac{|m_1\rangle_1 |m_2\rangle_2 - |m_2\rangle_1 |m_1\rangle_2}{\sqrt{2}}, \tag{10}$$

the action of $L_1^+ L_2^- + L_2^+ L_1^-$ on it can be given by,

$$\begin{aligned}&\left(L_1^\dagger L_2^- + L_2^\dagger L_1^-\right)|m_1, m_2\rangle \\ &= \frac{k_{1+}|m_1+1\rangle_1 k_{2-}|m_2-1\rangle_2 - k_{2+}|m_2+1\rangle_1 k_{1-}|m_1-1\rangle_2}{\sqrt{2}} + \frac{k_{1-}|m_1-1\rangle_1 k_{2+}|m_2+1\rangle_2 - k_{2-}|m_2-1\rangle_1 k_{1+}|m_1+1\rangle_2}{\sqrt{2}} \\ &= \frac{k_{1+}k_{2-}(|m_1+1\rangle_1|m_2-1\rangle_2 - |m_2-1\rangle_1|m_1+1\rangle_2)}{\sqrt{2}} + \frac{k_{1-}k_{2+}(|m_1-1\rangle_1|m_2+1\rangle_2 - |m_2+1\rangle_1|m_1-1\rangle_2)}{\sqrt{2}} \\ &= k_{1+}k_{2-}|m_1+1, m_2-1\rangle + k_{1-}k_{2+}|m_1-1, m_2+1\rangle. \end{aligned} \quad (11)$$

where the coefficients $k_{1+}$, $k_{1-}$, $k_{2+}$, and $k_{2-}$ are defined as,

$$\begin{aligned} k_{1+} &= \sqrt{s(s+1) - m_1(m_1+1)}, \\ k_{1-} &= \sqrt{s(s+1) - m_1(m_1-1)}, \\ k_{2+} &= \sqrt{s(s+1) - m_2(m_2+1)}, \\ k_{2-} &= \sqrt{s(s+1) - m_2(m_2-1)}. \end{aligned} \quad (12)$$

From (11), we can obtain,

$$\begin{aligned} L_1^+ L_2^- + L_2^+ L_1^- &= \sqrt{s(s+1) - m_1(m_1+1)}\sqrt{s(s+1) - m_2(m_2-1)}\hat{c}^\dagger_{m_1+1}\hat{c}_{m_1}\hat{c}^\dagger_{m_2-1}\hat{c}_{m_2}\hat{N}_{m_1}\hat{N}_{m_2} \\ &+ \sqrt{s(s+1) - m_1(m_1-1)}\sqrt{s(s+1) - m_2(m_2+1)}\hat{c}^\dagger_{m_1-1}\hat{c}_{m_1}\hat{c}^\dagger_{m_2+1}\hat{c}_{m_2}\hat{N}_{m_1}\hat{N}_{m_2}, \end{aligned} \quad (13)$$

where $\hat{c}^\dagger_m$ ($\hat{c}_m$) is the creation (annihilation) operator about the lowest Landau level $|s, m\rangle$.

It is noteworthy that there exists a special case when $m_1 - 1 = m_2$, i.e., $m_2 + 1 = m_1$. In this case, we have,

$$\begin{aligned} \left(L_1^+ L_2^- + L_2^+ L_1^-\right)|m_1, m_2\rangle &= \frac{\sqrt{s(s+1)-m_1(m_1+1)}\sqrt{s(s+1)-m_2(m_2-1)}(|m_1+1\rangle_1|m_2-1\rangle_2 - |m_2-1\rangle_1||m_1+1\rangle_2)}{\sqrt{2}} \\ &+ \frac{\sqrt{s(s+1)-m_1(m_1-1)}\sqrt{s(s+1)-m_2(m_2+1)}(|m_1-1\rangle_1|m_2+1\rangle_2 - |m_2+1\rangle_1,|m_1-1\rangle_2)}{\sqrt{2}} \\ &= \frac{\sqrt{s(s+1)-m_1(m_1+1)}\sqrt{s(s+1)-m_2(m_2-1)}(|m_1+1\rangle_1|m_2-1\rangle_2 - |m_2-1\rangle_1||m_1+1\rangle_2)}{\sqrt{2}} \\ &+ \frac{\sqrt{s(s+1)-m_1 m_2}\sqrt{s(s+1)-m_2 m_1}(|m_2\rangle_1|m_1\rangle_2 - |m_1\rangle_1|m_2\rangle_2)}{\sqrt{2}} \\ &= \sqrt{s(s+1)-m_1(m_1+1)}\sqrt{s(s+1)-m_2(m_2-1)}|m_1+1, m_2-1\rangle \\ &- \sqrt{s(s+1)-m_1 m_2}\sqrt{s(s+1)-m_2 m_1}|m_1, m_2\rangle. \end{aligned} \quad (14)$$

From (14), we can obtain,

$$\begin{aligned} \left(L_1^+ L_2^- + L_2^+ L_1^-\right)|m_1, m_2\rangle &= \sqrt{s(s+1)-m_1(m_1+1)}\sqrt{s(s+1)-m_2(m_2-1)}\hat{c}^\dagger_{m_1+1}\hat{c}_{m_1}\hat{c}^\dagger_{m_2-1}\hat{c}_{m_2}\hat{N}_{m_1}\hat{N}_{m_2} \\ &- \sqrt{s(s+1)-m_1 m_2}\sqrt{s(s+1)-m_2 m_1}\hat{N}_{m_1}\hat{N}_{m_2}. \end{aligned} \quad (15)$$

Substituting the (13) and (15) into (9), we can get the representation of the total squared angular momentum operator $L^2$ in the Landau levels basis with system size $N$,

$$\left(\sum_{m=-s}^{s} \overrightarrow{L_m}\right)^2 = \sum_{m=-s}^{s} s(s+1) \cdot \hat{N}_m + \sum_{m_i<m_j} \left(2m_i m_j \cdot \hat{N}_{m_i} \hat{N}_{m_j}\right)$$
$$+ \sum_{m_i<m_j} \left(\sqrt{s(s+1) - m_i(m_i+1)}\sqrt{s(s+1) - m_j(m_j-1)} \hat{c}^\dagger_{m_i+1} \hat{c}_{m_i} \hat{c}^\dagger_{m_j-1} \hat{c}_{m_j} \hat{N}_{m_i} \hat{N}_{m_j}\right)$$
$$- \sum_{m_i+1=m_j} \left(\sqrt{s(s+1) - m_i m_j}\sqrt{s(s+1) - m_j m_i} \hat{N}_{m_i} \hat{N}_{m_j}\right)$$
$$+ \sum_{m_i<m_j} \left(\sqrt{s(s+1) - m_i(m_i-1)}\sqrt{s(s+1) - m_j(m_j+1)} \hat{c}^\dagger_{m_i-1} \hat{c}_{m_i} \hat{c}^\dagger_{m_j+1} \hat{c}_{m_j} \hat{N}_{m_i} \hat{N}_{m_j}\right).$$
(16)


\fi

\end{document}